\documentclass[a4paper,10pt]{article}
\pdfoutput=1
\usepackage{fancyhdr}
\usepackage{changebar}
\usepackage{natbib}
\usepackage{graphicx}
\usepackage{amssymb,amsmath}
\usepackage{url}
\usepackage{color}
\usepackage{longtable}
\usepackage{subfig}
\usepackage{lscape}
\usepackage{rotating}
\usepackage{perpage}
\MakePerPage{footnote}
\usepackage[rmargin=4cm,lmargin=3cm,bmargin=4.5cm,tmargin=3.5cm]{geometry}
\usepackage{setspace}
\sffamily

\usepackage{graphicx}

\newcommand{\kph}{km\,s$^{-1}$}
\newcommand{\dg}{^{\circ}}
\newcommand{\m}{^{\rm m}}
\newcommand{\HI}{\textsc{Hi}}

\newcommand{\dmin}{\overset{\prime}{.}}
\newcommand{\dsec}{\overset{\prime\prime}{.\,}}
 
\bibpunct{(}{)}{,}{a}{}{;}

\title{{\bf Deep NIR Photometry of HI Galaxies\\Behind the Milky Way}}
\author{Wendy L. Williams$^{\rm 1}$, Patrick A. Woudt$^{\rm 1}$, Ren\'ee C. Kraan-Korteweg$^{\rm 1}$\\  \begin{small}Astrophysics, Cosmology and Gravity Centre, Astronomy Department,\end{small}\\ \begin{small} University of Cape Town, Private Bag X3, Rondebosch 7701, South Africa\end{small}\\  \begin{small}email: \texttt{wwilliams@strw.leidenuniv.nl}, \texttt{pwoudt@ast.uct.ac.za}, \texttt{kraan@ast.uct.ac.za}\end{small}}
\date{~}

\begin{document}

\maketitle

\begin{abstract}
\noindent
Current studies of the peculiar velocity flow field in the Local Universe are
limited by the lack of detection of galaxies behind the Milky Way. The
contribution of the largely unknown mass distribution in this ``Zone of
Avoidance'' (ZoA) to the dynamics of the Local group remains contraversial. We
have undertaken a near infrared (NIR) survey of \HI~detected galaxies in the
ZoA. The photomety derived here will be used in the NIR Tully-Fisher (TF)
relation to derive the peculiar velocities of this sample of galaxies in the
ZoA.
\end{abstract}

\section{Introduction}

The mass density field can be inferred from the peculiar velocities of galaxies,
independent of any a priori assumption on the relation or bias between visible
and dark matter. The determination of the peculiar velocity field requires
relatively large and uniform galaxy samples with high-fidelity distance
measurements independent of the observed redshift. This can for instance be
realized through the Tully-Fisher relation \citep[][]{TF1977}. However, peculiar
velocity surveys are hardly feasible in the ZoA where the obscuring effects of
dust and stars in the Milky Way prevent the identification of galaxies across
$10-20$\% of the sky. While this problem can be circumvented by statistical
interpolation of the mass distribution adjacent to the ZoA, various papers such
as those by \cite{Kolatt1995} and \cite{Loeb2008} suggest that these solutions
are inadequate and require unknown mass distributions to satisfactorily explain
the peculiar motion of the Local Group with respect to the Cosmic Microwave
Background. Several such dynamically important structures, including the Great
Attractor \citep[GA;][]{Lynden-Bell1988} and Local Void \citep{TullyFisher1987}
are known to reside within the ZoA. 

The 2MASS Tully-Fisher survey \citep[2MTF;][]{Masters+2008,Masters2008b} is  an
ongoing project that aims at measuring TF distances for all bright inclined
spirals in the 2MASS Redshift Survey  \citep[2MRS;][]{Huchra2005}.  The use of
the NIR bands will reduce the extinction effects due to the ZoA. Together with
the new high-fidelity \HI~measurements that are being obtained, the 2MTF will
provide better statistics for the study of the local peculiar velocity flow
field over the whole sky than ever before. However, there still remains a
significant part of the sky that will be excluded by the 2MTF. We aim to
provide TF data for galaxies in the ZoA which the 2MTF avoids.  These data will
be used to measure, for the \textit{first time}, the peculiar velocity flow
field \textit{within} the southern ZoA.

\section{HIZOA}
Blind \HI~surveys have been shown to be most effective at revealing galaxies in
the ZoA. The \HI~Parkes Deep Zone of Avoidance Survey, conducted  on the $64$~m
Parkes Radio Telescope, detected over $1000$ galaxies in the southern ZoA. With
an  exposure time five times longer than HIPASS, the average rms noise of the
survey was 6~mJy~beam$^{-1}$. It covered a velocity range of 
$-1200<v<12\,700$~\kph\ with a channel spacing of $13.2$~\kph. The survey 
covered the entire southern ZoA visible from Parkes: $212^{\circ} \le l \le
36^{\circ}$, $|b| < 5^{\circ}$  \citep{Henning2005};  $36^{\circ} < l <
52^{\circ}$ and $196^{\circ} < l< 212^{\circ}$, $|b| < 5^{\circ}$
\citep[Northern Extension;][]{Donley2005}; and $332^{\circ} < l < 36^{\circ}$,
$5^{\circ} < |b| < 10^{\circ}$ and $352^{\circ}<l<24^{\circ}$,
$10^{\circ}<|b|<15^{\circ}$  \citep[Galactic bulge extension;][]{Nebthesis}. 


The measurement of TF distances requires accurate \HI~linewidths, the fidelity
of which depends both on the velocity resolution and $S/N$ ratio of
the spectrum. Follow-up \HI~line observations were carried out for $82$ galaxies
with low $S/N$ and/or narrow linewidths in order to provide higher fidelity
linewidths for the TF analysis.

\section{NIR Follow-up Survey}
We have conducted a  dedicated follow-up NIR survey of HIZOA galaxies within
$6000$~\kph. The survey was conducted with the $1.4$~m IRSF telescope using the
SIRIUS camera for simultaneous imaging in the  $J$, $H$ and $K_s$ bands. The
survey images have an exposure time of $10$~min resulting in a limiting
magnitude approximately $2\m$ deeper than the 2MASS survey. The deeper images
and the superior resolution of the IRSF ($0\dsec45$~pix$^{-1}$) allow for the
detection of galaxies to higher levels of Galactic extinction and stellar
density. Moreover, the field-of-view of the IRSF ($7\dmin7 \times 7\dmin7$) is
ideally suited to this survey given the positional accuracy of the \HI~sources
of $\sim4'$ \citep{Donley2005}.

\subsection{Observations and Data Reduction}
The observations for the follow-up survey were started in 2006 and were
continued through to 2010. Unfortunately, a significant amount ($\sim60$\%) of
the $7$ weeks allocated exclusively to this project  was lost due to bad weather
and serious problems with the detector cooling system in 2010. The data were
reduced using the \texttt{SIRIUS} pipeline in IRAF which implements the standard
NIR data reduction, including dark current subtraction,  flat correction, sky
determination and subtraction, and frame to frame offset determination and
combination. The output images from the \texttt{SIRIUS} pipeline were
astrometrically and photometrically calibrated  using the 2MASS Point Source
Catalogue \citep[2MPSC][]{Strutskie2006} as a standard. The calibration was done
with a combination of IRAF and python scripts developed by Dr.~N.~Matsunaga and
modified by \cite{Riad2010}.


\subsection{Analysis}

Possible \HI~counterparts were identified in the respective NIR images by a
visual search of the three-colour images generated from the $K_s$ (red), $H$
(green) and $J$ (blue) bands.  Their different colour and extended nature allows
galaxies to be readily identified by eye.

\begin{figure}[t]
\centering
  \includegraphics[width=\textwidth]{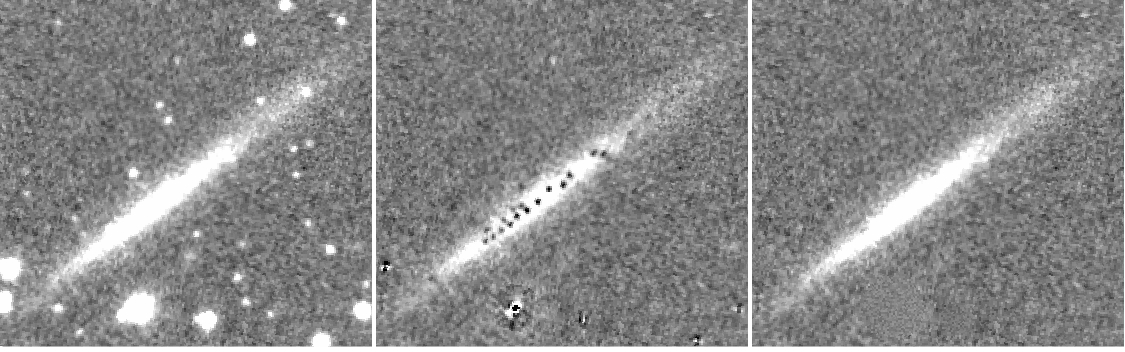}\\
  \includegraphics[width=\textwidth]{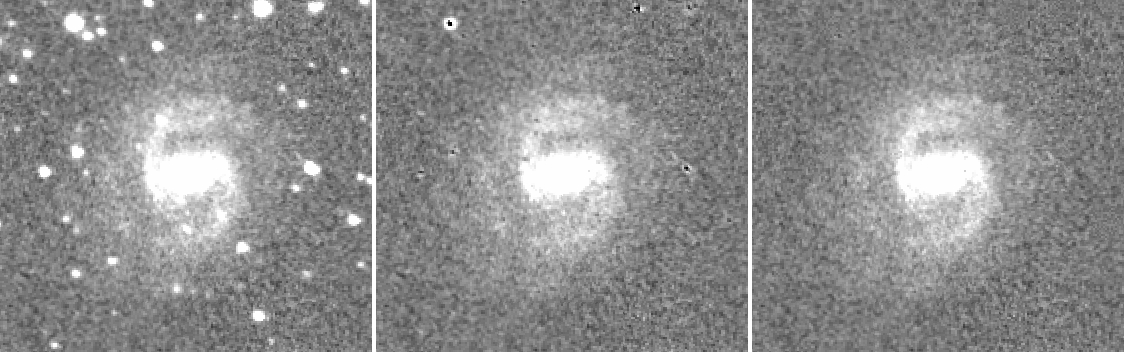}
 \caption[Improved performance of star-subtraction routine]{$K_s$ band images showing the improved performance of star-subtraction routine for two galaxies: J0716-18C (\textit{top} panel) and J0903-41 (\textit{bottom} panel). The \textit{left} panels show the original starry image, the \textit{middle} panel shows the star-subtraction using the original routine and the \textit{right} panels show the improved star-subtraction routine. The original routine leaves residuals on the galaxies which are not present with the new routines. Particularly affected are the planes of edge-on galaxies and substructure within galaxies. }
 \label{fig:ihabsub}
\end{figure}

The increase in stellar density near the Galactic plane results in heavy
contamination by foreground stars. Star-subtraction via PSF-fitting was employed
to remove the flux contribution of the foreground stars from the galaxy flux.
The automated PSF fitting routine for the Norma Wall Survey developed by Dr
Nagayama\footnote{Department of Astrophysics, Nagoya University} was  modified
to improve the subtraction of stars on edge-on disks and to prevent the removal
of sub-structure within the disks of face-on spirals. Figure~\ref{fig:ihabsub}
shows two examples of where the original star-subtraction routine resulted in
residuals on the galaxy.  A comparison of the results based on the original
star-subtraction method (central panel),  the newly developed method (right
panel) clearly demonstrates the improvement in the star-subtraction algorithm.

Isophotal elliptical aperture photometry was performed in each band, using the \texttt{ellipse} task in IRAF. The ellipticity and position angle of each galaxy was determined in each band and the  isophotal radii and magnitudes were calculated. A \textit{total} magnitude was determined by extrapolating an analytical double S\'{e}rsic function fitted to the outer parts of the disk \citep{Kirby+2008}.

\section{Early Results}

Of the $580$ fields that were observed, galaxies were found in $422$ fields. A
total of $567$ galaxies were identified given that $141$ fields had more than
one galaxy  identified as a possible counterpart.  Visual inspection of the  
\HI~spectra and NIR images of the $422$ \HI~galaxies for which possible NIR
counterparts were identified allowed for the confirmation of the counterpart for
$356$ ($84\%$) \HI~galaxies.  Of the $141$ fields with more than one
possible identified counterpart, a single counterpart could be confirmed for
$75$ fields, while the NIR counterpart remained ambiguous for $66$ fields.

\subsection{NIR Tully-Fisher Analysis}

\begin{sidewaysfigure}[h!]
\centering
 \includegraphics[width=\textwidth]{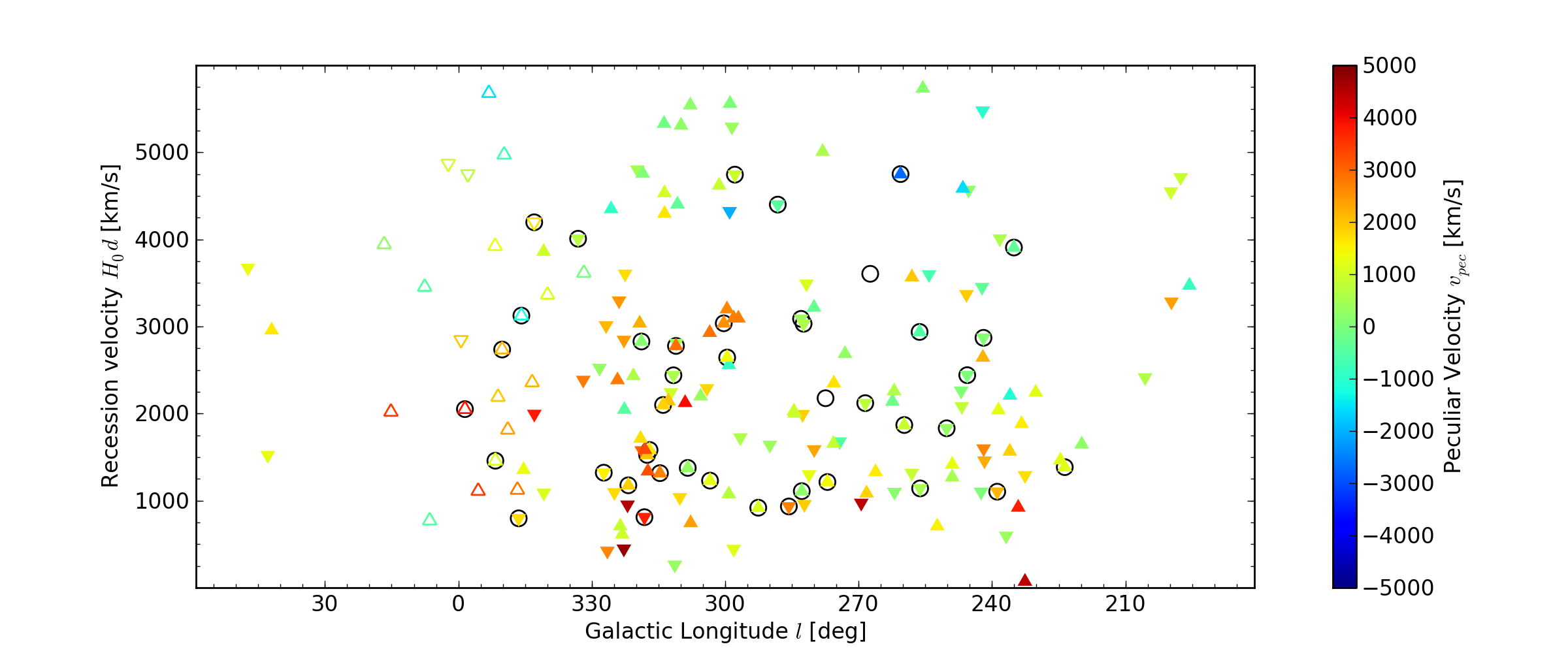}
 \caption[Peculiar velocities in the ZoA]{Peculiar velocities for all galaxies in the TF sample plotted in the Galactic longitude -- Hubble distance plane.  The colour scale shows the values of the peculiar velocities. The various symbols indicate the position of each galaxy with respect to the Galactic plane: upwards-pointing triangles lie above the plane ($0\dg < b< 5\dg$), downwards-pointing triangles lie below the plane ($-5\dg < b< 0\dg$). Near the Galactic bulge, in the GB extension of HIZOA, galaxies farther above the plane ($5\dg < b< 15\dg$) are plotted as open upwards-pointing triangles and those farther below the plane ($-10\dg < b< -5\dg$) are open downwards-pointing triangles. The points with large black circles around them have narrowband \HI~observations.}
 \label{fig:vpecfield}
\end{sidewaysfigure}

The peculiar velocities of $196$ edge-on galaxies with \emph{confirmed} NIR
cross-identifications of the \HI~source as well as good NIR photometry were
determined via the NIR TF relation \citep{Masters+2008}. This has led
to a preliminary determination of the peculiar velocity flow field in the
southern ZoA.  Figure~\ref{fig:vpecfield} shows the peculiar velocities for
these galaxies in the Galactic longitude -- Hubble distance plane.   Note that
the lack of galaxies with large positive peculiar velocities at larger distances
is due to the observational limit $v_{obs} = 6000$~\kph.  This initial map of
peculiar velocities indicates the large amplitude positive peculiar velocities
present in the foreground of the Great Attractor ($330\dg<l<270\dg$) at
$2000-3000$~\kph, compared to the more moderate and evenly distributed peculiar
velocities at similar distances in the longitude range $270\dg<l<210\dg$.

The derived peculiar velocity field, associated uncertainties and possible
systematic errors are discussed in more detail in \cite{WWThesis}.
Importantly, the extension of this NIR survey to slightly more distant galaxies,
$v_{obs} < 8000$~\kph~will allow the peculiar velocities behind the GA to be
mapped in more detail. This will better constrain the influence of the GA on the
motions of galaxies in the Local Universe. Finally, with the advent of more
sensitive \HI~and NIR instruments, the ability to extend this work to fainter,
less massive galaxies will become possible.\\~\\

\noindent {\bf Acknowledgements. ---}
\begin{small}
{We acknowledge the HIZOA survey team for early access to the data and
the organisers of this meeting. WW acknowledges the SA SKA project for funding
throughout this project. WW and RKK acknowledge financial support from the
National Research Foundation through a mobility grant from the South African -
Japanese (NRF/JSPS) bilateral grant in Astronomy.}
\end{small}

\begin{small}

\end{small}

\end{document}